\documentclass[12pt,a4paper,fleqn,twoside]{article}

\usepackage{amsmath}
\usepackage{amssymb}

\numberwithin{equation}{section} 
\newtheorem{theorem}{Theorem}[section]
\newtheorem{lemma}[theorem]{Lemma}
\newtheorem{definition}[theorem]{Definition}
\newtheorem{example}[theorem]{Example}
\newtheorem{proposition}[theorem]{Proposition}

\newenvironment{proof*}{\paragraph{Proof.}}{}

\renewcommand{\pmatrix}[2]{\left ( \begin{array}{#1} #2 \end{array} \right )}

\newcommand{\qed}{\square}

\newcommand{\arrow}{\rightarrow}
\newcommand{\map}{\mapsto}

\newcommand{\bb}[1]{\mathbb{#1}}

\newcommand{\Diff}[3]{\left . \frac{d}{d#1}#2\right |_{#3}}
\newcommand{\alg}{\mathfrak{g}}

\newcommand{\Ds}{\mathcal{D}}

\newcommand{\me}{\geqslant}
\newcommand{\les}{\leqslant}
\newcommand{\Dx}[1]{\partial_{x}^{#1}}

\newcommand{\deriv}[2]{\frac{\partial #1}{\partial #2}}
\newcommand{\bra}[1]{\left (#1\right )}

\newcommand{\pobr}[1]{\left \{#1\right \}}

\newcommand{\Matrix}[3]{\left (
\begin{array}{c}
#2\\ #3
\end{array}
\right )_{#1}}

\newcommand{\Matrixx}[4]{\left (
\begin{array}{c}
#2\\ #3\\ #4
\end{array}
\right )_{#1}}

\newcommand{\var}[2]{\frac{\delta #1}{\delta #2}}
\newcommand{\ad}{{\rm ad}}
\newcommand{\res}{{\rm res}}
\newcommand{\tr}{{\rm tr}}

\begin{document}

\title{Classical $R$-matrix theory of dispersionless systems: I. (1+1)-dimension theory}

\author{Maciej B\l aszak\footnote{E-mail: blaszakm@amu.edu.pl}$\ $ and B\l a\.zej M. Szablikowski\footnote{E-mail: bszablik@amu.edu.pl }\\Institute of Physics, A.Mickiewicz University,\\Umultowska 85, 61-614 Pozna\'n, Poland}



\maketitle

\begin{abstract}
A systematic way of construction of (1+1)-dimensional
dispersionless integrable Hamiltonian systems is presented. The
method is based on the classical R-matrix on Poisson algebras of
formal Laurent series. Results are illustrated with the known and
new (1+1)-dimensional dispersionless systems.\\ (To appear in J.
Phys. A: Math. Gen.)
\end{abstract}


\section{Introduction}

First order PDEs of the form
\begin{equation}
\frac{\partial u_{i}}{\partial
t}=\sum_{j=1}^{n}v_{ij}(u)\frac{\partial u_{j}}{\partial x},\qquad
i=1,...,n,  \label{1}
\end{equation}
are called \emph{hydrodynamic} or \emph{dispersionless} systems in
(1+1)-dimension. In this paper, we are interested in those PDEs
among (\ref{1}) which have multi-Hamiltonian structure, infinite
hierarchy of symmetries and conservation laws. The important
subclass of such systems are dispersionless limits of soliton
equations. Differential Poisson structures for hydrodynamic
systems were introduced for the first time by Dubrovin and Novikov
\cite{Du1} in the form
\begin{equation}
\pi _{ij}=g^{ij}(u)\partial _{x}-\sum_{k}\Gamma
_{k}^{ij}(u)\frac{\partial u_{k}}{\partial x},  \label{2}
\end{equation}
where $g^{ij}$ is a contravariant flat metric and $\Gamma
_{k}^{ij}$ are related coefficients of the contravariant
Levi-Civita connection. Then they were generalized by Mokhov and
Ferapontov \cite{Fe} to the nonlocal form
\begin{equation}
\pi _{ij}=g^{ij}(u)\partial _{x}-\sum_{k}\Gamma
_{k}^{ij}(u)\frac{\partial u_{k}}{\partial
x}+c\deriv{u_i}{x}\Dx{-1}\deriv{u_j}{x} \label{3a}
\end{equation}
in the case of constant curvature $c$. The natural geometric
setting of related bi-Hamiltonian structures (Poisson pencils) is
the theory of Frobenious manifolds based on the geometry of
pencils of contravariant metrics \cite{Du2}.

The other methods of the construction of dispersionless systems are
based on the application of the quasi-classical limit to the soliton theory.
For example, the quasi-classical limit of dressing method is considered by
Takasaki and Takebe \cite{TT}, while the quasi-classical limit of the
scalar nonlocal $\bar{\partial}$-problem is presented by Konopelchenko and Alonso \cite{KA}; see also the rich literature quoted in these papers.

In the following, we develop an alternative approach to construction
dispersionless systems and related Poisson pencils, based on an $R$-matrix theory. As it is well known, the $R$-matrix formalism proved very fruitful
in systematic construction of soliton systems (see for example
\cite{S-T-S}-\cite{Bl1} and the literature quoted therein). So, it seems
reasonable to develop such a formalism for dispersionless systems.
Recently, an important progress in
 that direction was made by Li \cite{Li} who applied the $R$-matrix theory to Poisson algebras \cite{G-KR}. In this paper, we apply his
results to a particular class of Poisson algebras.

The paper is organized as follows. In section 2 we briefly present
a number of basic facts and definitions concerning the formalism
applied. In section 3 we apply the formalism of the classical
$R$-matrix to the Poisson algebras of formal Laurent series. Then in
section 4 we illustrate our results with the known and new
(1+1)-dimensional integrable dispersionless systems.

\section{Hamiltonian dynamics on Lie algebras:\\$R$-Structures}

Let $\alg$ be a Lie algebra, $\alg^*$ the dual algebra related to
$\alg$ by the {\it duality map} $\langle \cdot,\cdot \rangle
\arrow \bb{R}$,
\begin{equation}
\alg^* \times \alg \arrow \bb{R}:\quad (\alpha,a)\map \langle
\alpha,a \rangle ,\qquad a\in \alg,\ \alpha \in \alg^*,
\end{equation}
and $\Ds(\alg^*):=\bb{C}^\infty(\alg^*)$ be a space of
$\bb{C}^\infty$-functions on $\alg^*$. Then, let
\begin{equation}
\ad:\alg \times \alg \arrow \alg:\quad (a,b)\map \ad_a b =
[a,b],\qquad a,b\in \alg,
\end{equation}
be adjoint action of $\alg$ on $\alg$, i.e. the Lie product, where
$[\cdot,\cdot]$ is a Lie bracket on $\alg$. There exists a natural
{\it Lie-Poisson bracket} on $\Ds(\alg^*)$. Let $F\in
\Ds(\alg^*)$, then a map $dF:\alg \arrow \alg$ such that
\begin{equation}\label{grad}
\Diff{t}{F(L+tL')}{t=0} = \langle L',dF(L) \rangle,\qquad L,L'\in
\alg^*,
\end{equation}
is a gradient of $F$. Let $L\in \alg^*$, functions $H,F$ belong to
the space of functions on $\alg^*:\ \Ds(\alg^*)$, and their
gradients $dH,dF\in \alg$, then the Lie-Poisson bracket reads
\begin{equation}\label{liepo}
\{H,F\}(L):=\langle L,[dF,dH] \rangle.
\end{equation}

We confine our further considerations to such algebras $\alg$ for
which its dual $\alg^*$ can be identified with $\alg$. So, we
assume the existence of a product $(\cdot,\cdot)_\alg$ on $\alg$
which is symmetric, non-degenerate and $\ad$-invariant:
\begin{equation}
(\ad_a b,c)_\alg + (b,\ad_a c)_\alg = 0, \qquad a,b,c\in \alg.
\end{equation}
Then, we can identify $\alg^*$ with  $\alg$, $(\alg^*\cong \alg)$
by setting
\begin{equation}
\langle \alpha,b \rangle = (a,b)_\alg,\qquad a,b\in \alg,\ \alpha
\in \alg^*,
\end{equation}
where $\alpha \in \alg^*$ is identified with $a\in \alg$. Now, we
can write the Lie-Poisson bracket as
\begin{align}
\{H,F\}(L)&=\langle L,[dF,dH] \rangle = (L,[dF,dH])_\alg \nonumber\\
&= (dF,[dH,L])_\alg =(dF,-ad_L dH)_\alg \equiv (dF,\theta(L)dH)_\alg,
\end{align}
where $\theta$ is a Poisson tensor $\theta:\alg \arrow \alg^*$.
Hence, the Hamiltonian dynamical system on $\alg^*$ can be defined
by the equation
\begin{equation}\label{hameq}
L_t = \theta(L) dH = -\ad_L dH = [dH,L].
\end{equation}
Now, we can identify the dynamic equation \eqref{hameq} and the
Lax equation with a natural Hamiltonian structure
\begin{equation}
L_t = [A,L] = \theta(L) dH = [dH,L].
\end{equation}

This abstract approach to integrable systems profits from a deeper
understanding of the nature of integrability as well as equips us
with a very general and efficient tool for the construction of
multi-Hamiltonian systems from scratch.

\begin{definition}
An $R$-structure is a Lie algebra $\alg$ equipped with a linear map\\
$R:\alg \arrow \alg$ (called the classical $R$-matrix) such that
the bracket
\begin{equation}\label{lieR}
[a,b]_R:= [Ra,b]+[a,Rb],\qquad a,b\in \alg,
\end{equation}
is a second Lie product on $\alg$.
\end{definition}

\begin{definition}
Let $A$ be a commutative, associative algebra with unit $1$. If
there is a Lie bracket on $A$ such that for each element $a\in A$,
the operator $\ad_a:b\map [a,b]$ is a derivation of the
multiplication, then $(A,[.,.])$ is called a \rm{Poisson
algebra}.
\end{definition}
Thus the Poisson algebras are Lie algebras with an additional
associative algebra structure (with commutative multiplication and
unit $1$) related by the derivation property to the Lie bracket.

\begin{theorem} \cite{Li}
Let $A$ be a Poisson algebra with Lie bracket $[\cdot,\cdot]$ and
non-degenerate ad-invariant pairing $(\cdot,\cdot)_A$ with respect
to which the operation of multiplication is symmetric, i.e.
$(ab,c)_A=(a,bc)_A,\ \forall a,b,c\in A$. Assume $R\in {\rm End}(A)$ is
a classical $R$-matrix, then for each integer $n\me -1$, the
formula
\begin{equation}\label{pobr}
\{H,F\}_n = (L,[R(L^{n+1}dF),dH]+[dF,R(L^{n+1}dH)])_A,
\end{equation}
where $H,F$ are smooth functions on $A$, defines a Poisson
structure on $A$. Moreover, all $\{\cdot,\cdot\}_n$ are
compatible.
\end{theorem}
The related Poisson bivectors $\pi_n$ are given by the following
Poisson maps
\begin{equation}\label{pob}
\pi_n :dH\map -\ad_L R(L^{n+1}dH) - L^{n+1} R^* (\ad_L dH),\qquad
n\me -1,
\end{equation}
where the adjoint of $R$ is defined by the relation
\begin{equation}
(a,Rb)_A=(R^*a,b)_A.
\end{equation}
Notice that the bracket \eqref{pobr} with $n=-1$ is just a
Lie-Poisson bracket with respect to Lie bracket \eqref{lieR}
\begin{equation}\label{liepor}
\{H,F\}_{-1} = (L,[dF,dH]_R)_A.
\end{equation}

We will look for a natural set of functions in involution w.r.t.
the Poisson brackets \eqref{pobr}. A smooth function $F$ on $A$ is
ad-invariant if $dF\in \ker ad_L$, i.e
\begin{equation}
[dF,L]=0,\qquad L\in A,
\end{equation}
which are Casimir functionals of the natural Lie-Poisson bracket
\eqref{liepo}.

Hence, the following Lemma is valid
\begin{lemma}\cite{Li}
 Smooth functions on $A$ which are ad-invariant commute in $\{\cdot,\cdot\}_n$. The Hamiltonian system generated by a smooth ad-invariant function $C(L)$ and the Poisson structure $\{\cdot,\cdot\}_n$ is given by the Lax equation
\begin{equation}
L_t = [R(L^{n+1}dC),L],\qquad L\in A.
\end{equation}
\end{lemma}

Let us assume that an appropriate product on Poisson algebra $A$ is given by {\it the trace form} $\tr :A \arrow \mathbb{R}$
\begin{equation}
(a,b)_A = \tr(ab),\qquad a,b\in A.
\end{equation}
As we have assumed a nondegenerate trace form $\tr$ on $A$, we will
consider the most natural Casimir functionals given by the trace
of powers of $L$, i.e.
\begin{equation}\label{casq}
C_q(L) = \frac{1}{q+1} \tr(L^{q+1}).
\end{equation}
The related gradients by \eqref{grad} are of the form
\begin{equation}
dC_q(L) = L^q.
\end{equation}
Then taking these $C_q(L)$ as Hamiltonian functions, one finds a
hierarchy of evolution equations which are multi-Hamiltonian
dynamical systems
\begin{eqnarray}\label{eveq}
L_{t_q} = [R(dC_q),L]= \pi_{-1}(dC_q) = \pi_{0}(dC_{q-1}) = ... =\pi_{l}(dC_{q-l-1}) = ...\ .
\end{eqnarray}

For any $R$-matrix each two evolution equations in the hierarchy
\eqref{eveq} commute due to the involutivity of the Casimir
functions $C_q$. Each equation admits all the Casimir functions as
a set of conserved quantities in involution. In this sense we will
regard \eqref{eveq} as a hierarchy of \emph{integrable} evolution
equations.

To construct the simplest $R$-structure let us assume that the
Poisson algebra $A$ can be split into a direct sum of Lie
subalgebras $A_+$ and $A_-$, i.e.
\begin{equation}
A =A_+ \oplus A_-,\qquad [A_\pm,A_\pm]\subset A_\pm.
\end{equation}
Denoting the projections onto these subalgebras by $P_\pm$, we
define the $R$-matrix as
\begin{equation}\label{rp}
R = \frac{1}{2} (P_+ - P_-)
\end{equation}
which is well defined.

 Following the above scheme, we are able to construct in a systematic way integrable multi-Hamiltonian dispersionless systems, with infinite hierarchy of involutive constants of motion and infinite hierarchy of related commuting symmetries, ones we fix a Poisson algebra.

\section{Poisson algebras of formal Laurent series}

 Let $A$ be an algebra of Laurent series with respect to $p$ \cite{Bl2}
\begin{equation}
A =\left \{L = \sum_{i\in \bb{Z}} u_i(x)p^i\right \},
\end{equation}
where the coefficients $u_i(x)$ are smooth functions. It is
obviously commutative and associative algebra under
multiplication. The Lie-bracket can be introduced in infinitely
many ways as
\begin{equation}\label{liebr}
[L_1,L_2]= p^r \bra{\deriv{L_1}{p} \deriv{L_2}{x} - \deriv{L_1}{x}
\deriv{L_2}{p}}:=\{L_1,L_2\}_r,\qquad r\in \bb{Z},
\end{equation}
as $\ad_L = p^r (\deriv{L}{p} \deriv{}{x} - \deriv{L}{x}
\deriv{}{p})$ is a derivation of the multiplication, so
$A_r:=(A,\{\cdot,\cdot\}_r)$ are Poisson algebras.
\begin{lemma}
An appropriate symmetric product on $A_r$ is given by a trace form
$(a,b)_A:=\tr(ab)$:
\begin{equation}\label{trace}
\tr L = \int_\Omega \res_r Ldx,\qquad \res_r L = u_{r-1}(x),
\end{equation}
which is $\ad$-invariant. In expression \eqref{trace} the integration
denotes the equivalence class of differential expressions modulo
total derivatives.
\end{lemma}
\begin{proof*}
We assume that $\Omega =\bb{S}^1$ if $u$ is periodic or $\Omega =
\bb{R}$ if $u$ belong to the Schwartz space. The Symmetry is
obvious as $L_1L_2=L_2L_1$. Let $L_1,L_2\in A:$ \
$L_1=\sum_{i}a_{i}p^{i},\;L_2=\sum_{j}b_{j}p^{j}$, then
\begin{eqnarray}\label{trres}
res_{r}[L_1,L_2]=
res_{r}\bra{p^{r}\sum_{i,j}(ia_{i}(b_j)_x-j(a_i)_xb_{j})p^{i+j-1}}
=\sum_{i}i(a_{i}b_{-i})_{x}.
\end{eqnarray}
So, $\tr[L_1,L_2]=0$ and hence
\begin{align}
&tr([A,B]C)+tr(B[A,C])\nonumber \\
&=tr([A,BC]-B[A,C])+tr(B[A,C])=tr[A,BC]=0.\qquad \qed \nonumber
\end{align}
\end{proof*}

For a given functional $F(L)= \int_\Omega f(u)dx$, we define its
gradient as
\begin{equation}\label{res}
dF = \var{F}{L} = \sum_i \var{f}{u_i}p^{r-1-i},
\end{equation}
where $\delta f/\delta u_i$ is a variational derivative.

We construct the simplest $R$-matrix, through a decomposition of
$A$ into a direct sum of Lie subalgebras. For a fixed $r$ let
\begin{equation}
\begin{split}
A_{\me -r+k}= P_{\me -r+k}A   =\left \{L = \sum_{i\me -r+k} u_i(x)p^i\right \},\\
A_{< -r+k}= P_{< -r+k}A   =\left \{L = \sum_{i< -r+k}
u_i(x)p^i\right \},
\end{split}
\end{equation}
where $P$ are appropriate projections.

\begin{proposition}
$A_{\me -r+k}, A_{< -r+k}$ are Lie subalgebras in the following
cases:
\begin{enumerate}
\item[1.] $\quad k=0,\ r=0$,
\item[2.] $\quad k=1,2,\ r\in \bb{Z}$.
\end{enumerate}
\end{proposition}
The proof is through a simple inspection. Then, the $R$-matrix is
given by the projections
\begin{equation}\label{rmat}
R = \frac{1}{2}(P_{\me -r+k} - P_{< -r+k}) = P_{\me -r+k} -
\frac{1}{2} = \frac{1}{2} - P_{< -r+k}.
\end{equation}
To find $R^*$ one has to find $P_{\me -r+k}^*$ and $P_{< -r+k}^*$
given by the orthogonality relations
\begin{equation}
(P_{\me -r+k}^*,P_{< -r+k}) = (P_{< -r+k}^*,P_{\me -r+k})=0.
\end{equation}
So, we have
\begin{equation}
P_{\me -r+k}^* = P_{< 2r-k},\qquad P_{< -r+k}^* = P_{\me 2r-k},
\end{equation}
and then
\begin{equation}
R^* = \frac{1}{2}(P_{\me -r+k}^* - P_{< -r+k}^*) = \frac{1}{2} -
P_{\me 2r-k} = P_{< 2r-k} - \frac{1}{2}.
\end{equation}

Hence, the hierarchy of evolution equations \eqref{eveq} for
Casimir functionals $C(L)$ with $R$-matrix given by \eqref{rmat}
has the form of two equivalent representations
\begin{equation}\label{laxh}
L_{t_q} = \{(L^q)_{\me -r+k},L\}_r = -\{(L^q)_{<
-r+k},L\}_r,\qquad L\in A,
\end{equation}
which are Lax hierarchies.

We have to explain what type of Lax operators can be used in
(\ref{laxh}) to obtain a consistent operator evolution equivalent
with some nonlinear integrable equation. Here, we are interested
in extracting closed systems for a finite number of fields. The
case of infinite number of fields was considered recently in
\cite{Bl2}. Hence, we start with looking for Lax operators L in
the general form
\begin{equation}\label{laxo}
L = u_N p^N + u_{N-1} p^{N-1} + ... + u_{-m+1} p^{-m+1} + u_{-m}
p^{-m}
\end{equation}
of N-th order, parametrized by finite number of fields $u_i$. To
obtain a consistent Lax equation, the Lax operator (\ref{laxo})
has to form a proper submanifold of the full Poisson algebra under
consideration, i.e. the left and right-hand sides of expression
(\ref{laxh}) have to lie inside this submanifold.

Observing (\ref{laxh}) with some $(L^q)_{< -r+k}=
a_{-r+k-1}p^{-r+k-1}+a_{-r+k-2}p^{-r+k-2}+...$ one immediately
obtains the highest order of the right-hand side of Lax equation
as
\begin{align}\label{high}
L_{t} &= (u_N)_t p^N + (u_{N-1})_t p^{N-1} + ...\nonumber\\
&= -\{(L^q)_{< -r+k},u_N p^N + lower\}_r\nonumber\\
&= -((-r+k-1)a_{-r+k-1}(u_N)_x - N(a_{-r+k-1})_x u_N) p^{N+k-2}+
lower,
\end{align}
where $lower$ represents lower orders. Observing (\ref{laxh}) with
some $\ (L^q)_{\me -r+k} = ... + a_{-r+k+1}p^{-r+k+1} +
a_{-r+k}p^{-r+k}$ one immediately obtains the lowest order of the
right-hand side of Lax equation (\ref{laxh}) as
\begin{align}\label{low}
L_{t} &= ... + (u_{-m+1})_t p^{-m+1} + (u_{-m})_t p^{-m}\nonumber\\
& = \{(L^q)_{\me -r+k}, higher + u_{-m} p^{-m}\}_r\nonumber\\
&= higher +((-r+k)a_{-r+k}(u_{-m})_x - (-m)(a_{-r+k})_x u_{-m})
p^{-m+k-1},
\end{align}
where $higher$ represents higher orders. Simple consideration of
\eqref{high} and \eqref{low} with the condition $N\me -m$ leads to
the admissible Lax polynomials with a finite number of field
coordinates, which form proper submanifolds of Poisson
subalgebras. They are given in the form
\begin{eqnarray}
\label{laxk0} k=0,\ r=0\ :&\quad L = c_N p^N + c_{N-1} p^{N-1} + u_{N-2}
p^{N-2} + ... + u_1 p + u_0,\\
\label{laxk1} k=1,\ r\in \bb{Z}\
:&\quad L = c_N p^N + u_{N-1} p^{N-1} + ... + u_{1-m} p^{1-m} +
u_{-m} p^{-m},\\
\label{laxk2} k=2,\ r\in \bb{Z}\ :&\quad L = u_N
p^N + u_{N-1} p^{N-1} + ... + u_{1-m} p^{1-m} + c_{-m} p^{-m},
\end{eqnarray}
where the $u_i$ are dynamical fields and $c_N,c_{N-1},c_{-m}$ are
arbitrary time independent functions of $x$.

Once we know the restricted Lax operators $L\in A$, we can now
investigate the form of gradients of Casimir functionals given by
powers of $L$, as well as we can investigate some further
simplest admissible reductions of Lax operators.

In general, the fractional powers of $L$ will lead to interesting
results. Let $L$ be given by \eqref{laxo}, then we consider
polynomials of the form
\begin{equation}
\begin{split}
&L^{\frac{1}{N}} = a_1 p + a_0 + a_{-1} p^{-1} + ..., &\qquad \mbox{ for } N\in \mathbb{Z}_{+},\\
&L^{\frac{1}{N}} = a_{-1} p^{-1} + a_{-2}p^{-2} + a_{-3} p^{-3} + ..., &\qquad \mbox{ for } N\in \mathbb{Z}_{-},\\
&L^{\frac{1}{m}} = ... + b_1 p + b_0 + b_{-1} p^{-1}, &\qquad \mbox{ for } m\in \mathbb{Z}_{+},\\
&L^{\frac{1}{m}} = ... + b_3 p^3 + b_2 p^2 + b_1 p, &\qquad \mbox{ for } m\in \mathbb{Z}_{-},
\end{split}
\end{equation}
where the coefficients $a_i$ and $b_i$ are obtained by requiring
$(L^{\frac{1}{N}})^N=L$ and $(L^{\frac{1}{m}})^m=L$, successively
via the recurrent procedure. Therefore, one finds the formal
expansion of $L^{\frac{1}{N}}$ and $L^{\frac{1}{m}}$ and one can
calculate the fractional powers of $L$ for integer $q$:
$L^{\frac{q}{N}}$ and $L^{\frac{q}{m}}$. Notice that they are in
the form of infinite series, except the case of integer powers,
obviously. In fact, we need only their finite parts
$(L^{\frac{q}{N}})_{\me -r+k}$ or $(L^{\frac{q}{m}})_{< -r+k}$.
Hence, for a given $L\in A$ in principle we can construct two
different hierarchies of Lax equations \eqref{laxh}.

\paragraph{The case of $k=0$.}
Let us consider Lax operators of the form \eqref{laxk0}. One can
see that $L^{\frac{q}{N}}$ has the form
\begin{equation}\label{gck0}
L^{\frac{q}{N}} = \alpha_q p^q + \alpha_{q-1} p^{q-1} + a_{q-2}
p^{q-2} + a_{q-3} p^{q-3} + lower,\qquad q\in \mathbb{Z}_+,
\end{equation}
where $\alpha_i,\alpha_{i-1}$ are arbitrary $x$ independent
functions. The second form $L^{\frac{1}{m}}$, since $m=0$, gives
only the integer powers of $L$, such that $(L^q)_{\me 0} = L^q$,
leading to trivial dynamics $L_t = \{L^q,L\}_0 = 0$. Hence, for
$k=0$ there is only one Lax hierarchy for gradients of Casimir
functionals \eqref{gck0}. There are no further reductions.

\paragraph{The case of $k=1$.}
Let us consider Lax operators of the form \eqref{laxk1}. One can
see that $L^{\frac{q}{N}}$ and $L^{\frac{q}{m}}$ have the
forms
\begin{align}
\label{gck1a} &L^{\frac{q}{N}} = \alpha_q p^q + a_{q-1} p^{q-1} + a_{q-2}
p^{q-2} + a_{q-3} p^{q-3} + + lower &\quad q\in \mathbb{Z}_+,\\
\label{gck1b} &L^{\frac{q}{m}} = higher + a_{3-q}p^{3-q} + a_{2-q} p^{2-q} + a_{1-q} p^{1-q} + u_{-m}^{\frac{q}{m}}p^{-q} &\quad q\in \mathbb{Z}_+,
\end{align}
where $\alpha_i$ is an arbitrary $x$-independent function. We
remark that there is always a further admissible reduction of
equations \eqref{laxh} given by $u_{-m}=0$, since such reduced Lax
polynomial would still be of the form \eqref{laxk1}. After such
reduction we have to look for the form of gradients of Casimir
functionals. By putting $u_{-m}=0$ in \eqref{gck1a}, it preserves
the order of highest terms and the form. For \eqref{gck1b} the
lowest order disappears, and as all other terms depend linearly on
the powers of $u_{-m}$, such an $L^{\frac{q}{m}}$ will reduce to
zero, except the one case for $q=m$. We can continue the
reductions by putting $u_{1-m}=0$ and so on. Therefore, the
reductions are proper in general only for the gradients of Casimir
functionals in the form \eqref{gck1a}.

\paragraph{The case of $k=2$.}
Let us consider Lax operators of the form \eqref{laxk2}. One can
see that $L^{\frac{q}{N}}$ and $L^{\frac{q}{m}}$ take the
form
\begin{align}
\label{gck2a} &L^{\frac{q}{N}} = u^{\frac{q}{N}}_q p^q + a_{q-1} p^{q-1} + a_{q-2} p^{q-2} +
a_{q-3} p^{q-3} + lower &\quad q\in \mathbb{Z}_+,\\
\label{gck2b} &L^{\frac{q}{m}} = higher + a_{3-q}p^{3-q} + a_{2-q} p^{2-q} + a_{1-q} p^{1-q} + \alpha_{-m} p^{-q} &\quad q\in \mathbb{Z}_+,
\end{align}
where $\alpha_i$ is an arbitrary $x$-independent function. We
remark that there is always a further admissible reduction of
equations \eqref{laxh} given by $u_{N}=0$, since such a reduced
Lax polynomials would still be of the form \eqref{laxk2}. The next
reduction is $u_{N-1}=0$ and so on. By analogous considerations as
for $k=1$, these reductions are proper in general only for the
gradients of Casimir functionals in the form \eqref{gck2b}.

The different schemes are interrelated as it is explained in the
following theorem.
\begin{theorem}\label{rel}
Under the transformation
\begin{equation}\label{transf}
x'=x,\ p'=p^{-1},\ t'=t
\end{equation}
the Lax hierarchy defined by $k=1,\ r$ and $L$ transforms into the
Lax hierarchy defined by $k=2,\ r'=2-r$ and $L'$, i.e.
\begin{equation}
k=1,r,L\Longleftrightarrow k=2,r'=2-r,L'.
\end{equation}
\end{theorem}
\begin{proof*}
It is readily seen that the Lax operators for $k=1$ and $r$ of the
forms \eqref{laxk1} transform into the well restricted Lax
operators for $k=2$ and $r'=2-r$ of the forms \eqref{laxk2}. Let's
observe that
\[ \{A,B\}_r = p^r (\deriv{A}{p}
\deriv{B}{x} - \deriv{A}{x} \deriv{B}{p}) = -p'^{-r+2}
(\deriv{A'}{p'} \deriv{B'}{x'} - \deriv{A'}{x'} \deriv{B'}{p'}) =
-\{A',B'\}'_{r'},\]
and
\[ (dC)_{\me s}' = (dC')_{\les -s}. \]
Hence, we have
\begin{align*}
L_t &= \{(dC)_{\me -r+1},L\}_r = -\{(dC)'_{\me -r+1},L'\}'_{r'}\\
& =  -\{(dC')_{\les r-1},L'\}'_{r'} =  -\{(dC')_{<
-r'+2},L'\}'_{r'} = L'_{t'}.\qquad \qed
\end{align*}
\end{proof*}

Therefore, some dispersionless systems can be reconstructed from
different Poisson algebras. Moreover, we remark that the gradients
of Casimir functionals for $k=1$ of the form \eqref{gck1a},
\eqref{gck1b} by $p'=p^{-1}$ transform into \eqref{gck2b},
\eqref{gck2a} for $k=2$, respectively, at a slant.

Two equivalent representations of Poisson bivectors \eqref{pob}
with the $R$-matrix given by \eqref{rmat} are defined through the
following Poisson maps
\begin{align}
\label{poten}
\pi_n dH &= \{(L^{n+1}dH)_{\me -r+k},L\}_r +L^{n+1} (\{L,dH\}_r)_{\me 2r-k}\nonumber\\
&= -\{(L^{n+1}dH)_{< -r+k},L\}_r - L^{n+1}(\{L,dH\}_r)_{<2r-k},\qquad n\me -1.
\end{align}
It turns out that the first representation yields a direct access
to the lowest polynomial order of $\pi_n dH$, whereas the second
representation yields information about the highest orders
present. There are two options. The best situation is when a given
Lax operator forms a proper submanifold of the full Poisson
algebra, i.e. the image of the Poisson operator $\pi_n$ lies in
the space tangent to this submanifold for each element. If this is
not the case, {\it the Dirac reduction} can be invoked for
restriction of a given Poisson tensor to a suitable submanifold.

\paragraph{The case of $k=0$.}
Let us consider the simplest admissible Lax polynomial
\eqref{laxk0} of the form
\begin{equation}\label{slaxk0}
L = p^N + u_{N-2} p^{N-2} + ... + u_1 p + u_0.
\end{equation}
This is the well-known dispersionless Gelfand-Dickey case. Then,
the gradient of the functional $H(L)$ is given in the form
\begin{equation}
\var{H}{L} = \var{H}{u_0} p^{-1} + \var{H}{u_1} p^{-2} + ... +
\var{H}{u_{N-2}} p^{1-N}.
\end{equation}

Observing \eqref{poten} for $n=-1$ one immediately obtains the
highest and lowest order of $\pi_{-1}dH$ as
\begin{equation}
\pi_{-1}\bra{\var{H}{L}} = \alpha_{N-2} p^{N-2} + \alpha_{N-3}
p^{N-3} + ... + \alpha_1 p + \alpha_0.
\end{equation}
Hence $\pi_{-1}dH$ is tangent to the submanifold formed by the Lax
operator of the form \eqref{slaxk0}, and the linear Poisson
structure, since $\bra{\var{H}{L}}_{\me 0} = 0$, is given by
\begin{equation}\label{link0}
\pi_{-1} \bra{\var{H}{L}} = \bra{\pobr{L,\var{H}{L}}_0}_{\me 0}.
\end{equation}

For $n=0$, $L$ does not define a proper Poisson submanifold, as
\[ \pi _{0}\bra{\var{H}{L}} = \alpha_{N-1} p^{N-1} + \alpha_{N-2}
p^{N-2} + ... + \alpha_1 p + \alpha_0,\]
and a Dirac reduction is required. Let
\begin{equation}
\overline{L}=p^{N}+up^{N-1}+u_{N-2}p^{N-2}+...+u_{0}=L+up^{N-1}
\end{equation}
be a an extended Lax polynomial and we shall consider the $\pi
_{0}$ Hamiltonian flow for $\overline{L}$ together with the
constraint $u=0$. However, imposition of such a constraint leads
to the modification of the $\pi_{0}$ Poisson structure due to the
Dirac reduction. We briefly remind the calculation procedure on
the example considered. The Hamiltonian flow for $u$, given by the
coefficient of $p^{N-1}$ in the Hamiltonian equation for
$\overline{L}$, under the constraint $u=0$, gives the relation
\begin{equation}\label{ut}
\left . u_{t} \right |_{u=0} =
\bra{res_0\pobr{\var{H}{\overline{L}},\overline{L}}_0}_{u=0} = 0,
\end{equation}
where
\begin{equation}
\var{H}{\overline{L}}=\var{H}{L}+\var{H}{u}p^{-N}.
\end{equation}
Then solving (\ref{ut}) with respect to $\var{H}{u}$ one gets
\begin{equation}
\frac{\delta H}{\delta u}=-\frac{1}{N}\partial _{x}^{-1}res_{0}\left\{ L,%
\frac{\delta H}{\delta L}\right\} _{0}.
\end{equation}
It means that the function $\frac{\delta H}{\delta u}$ can be
expressed in terms of $\frac{\delta H}{\delta u_{i}}$. This
implies
\begin{align}
\label{quadk0}
&\pi_{0}^{red}\bra{\var{H}{L}}\equiv \pi_{0}\bra{\var{H}{\overline{L}}}_{u=0}
=\left\{ \left( L\frac{\delta H}{\delta \overline{L}}\right)_{\me 0},L\right\}_{0}+L\left( \left\{ L,\frac{\delta H}{\delta \overline{L}}\right\}_{0}\right)_{\me 0}\nonumber\\
&=\left\{ \left( L\frac{\delta H}{\delta L}+L\frac{\delta H}{\delta u}p^{-N}\right)_{\me 0},L\right\}_{0}+L\left( \left\{ L,\frac{\delta H}{\delta L}+\frac{\delta H}{\delta u}p^{-N}\right\} _{0}\right)_{\me 0}\nonumber\\
&=\left\{ \left( L\frac{\delta H}{\delta L}\right)_{\me 0},L\right\}_{0}+\left\{ \frac{\delta H}{\delta u},L\right\} _{0}+L\left( \left\{ L,\frac{\delta H}{\delta L}\right\}_{0}\right)_{\me 0}\nonumber\\
&=\left\{ \left( L\frac{\delta H}{\delta L}\right)_{\me
0},L\right\}_{0}+L\left( \left\{ L,\frac{\delta H}{\delta
L}\right\}_{0}\right)_{\me 0}+\frac{1}{N}\left\{
L,\partial_{x}^{-1}res_{0}\left\{L,\frac{\delta H}{\delta
L}\right\}_{0}\right\}_{0},
\end{align}
i.e. the second Poisson map of dispersionless Gelfand-Dickey
systems. Poisson structures $\pi _{-1}$ and $\pi_{0}^{red}$ were
constructed for the first time in \cite{F-OR} as the
dispersionless limit of the Poisson structures of the
Gelfand-Dickey soliton systems. Notice that $\pi_{0}^{red}$ is
purely differential due to the property \eqref{trres}.

Observing \eqref{poten} for $n\me 1$ one obtains the highest and
lowest order of $\pi_{n}dH$ as
\begin{equation}
\pi_{n}\bra{\var{H}{L}} = \alpha_{(n+1)N-1} p^{(n+1)N-1} +
\alpha_{(n+1)N-2} p^{(n+1)N-2} + ... + \alpha_1 p + \alpha_0.
\end{equation}
Hence, the polynomials of the form \eqref{slaxk0} do not form a
proper Poisson submanifold. In fact there is not obvious proper
Poisson submanifold for $\pi_n$ with $n\me 1$, apart from the
trivial case of the first order polynomials with $n=1$.
Nevertheless, the Dirac reduction can be invoked to restrict the
bivectors $\pi_n$ on the polynomials to the submanifold of the
form \eqref{poten}.

\paragraph{The case of $k=1$.}
This case contains new results. Let us consider the simplest
admissible Lax polynomial \eqref{laxk1} of the form
\begin{equation}\label{slaxk1}
L = p^N + u_{N-1} p^{N-1} + ... + u_{1-m} p^{1-m} + u_{-m} p^{-m}.
\end{equation}
Then gradient of the functional $H(L)$ is given in the form
\begin{equation}
\var{H}{L} = \var{H}{u_{-m}} p^{r+m-1} + \var{H}{u_{-m+1}}
p^{r+m-2} + ... + \var{H}{u_{N-1}} p^{r-N}.
\end{equation}

Observing \eqref{poten} for $n=-1$ one obtains the highest and
lowest order of $\pi_{-1}\bra{\var{H}{L}}$ as
\begin{align*}
\pi_{-1}\bra{\var{H}{L}} &= \bra{(...)p^{N-1} + lower} + \bra{(...)p^{2r-2} + lower}\\
&= \bra{higher + (...)p^{-m}} + \bra{higher + (...)p^{2r-1}},
\end{align*}
where $lower$ ($higher$) represents lower (higher) orders. Hence,
the Lax operators of the type \eqref{slaxk1} form a proper
submanifold for $N\me 2r-1\me -m$, as then
$\pi_{-1}\bra{\var{H}{L}}$ is tangent to this submanifold. So the
linear Poisson map is
\begin{equation}\label{link1}
\pi_{-1}\bra{\var{H}{L}} = \pobr{\bra{\var{H}{L}}_{\me
-r+1},L}_r+\bra{\pobr{L,\var{H}{L}}_r}_{\me 2r-1}.
\end{equation}
Otherwise a Dirac reduction is required.

For the second Poisson map with $n=0$, $L$ does not define a
proper Poisson submanifold and two distinct cases have to be
considered.
\paragraph{$2r-1\me 1:$}
\begin{equation*}
\pi_{0}\left( \frac{\delta H}{\delta L}\right) = (...)
p^{(N-1)+(2r-1)}+...+(...)p^{N-1}+...+(...)p^{-m},
\end{equation*}
hence $L$ is not properly defined and a Dirac reduction is
required for additional higher order terms. The simplest case is
$r=1$ with one-field reduction. Let
\[ \overline{L}=up^{N}+u_{N-1}p^{N-1}+u_{N-1}p^{N-2}+...+u_{1-m}p^{1-m}+u_{-m}p^{-m}.\]
The Dirac reduction with the constraint $u=1$ leads to the second
Poisson map in the form
\begin{align}
\label{quadk1a}
&\pi _{0}^{red}\left( \frac{\delta H}{\delta L}\right)=\nonumber\\
&\left\{ \left( L\frac{\delta H}{\delta L}\right)_{\geq
0},L\right\}_{1}+L\left(\left\{L,\frac{\delta H}{\delta
L}\right\}_{1}\right)_{\geq
1}+\frac{1}{N}\left\{L,\partial_{x}^{-1}res_{1}\left\{L,\frac{\delta
H}{\delta L}\right\}_{1}\right\}_{1},
\end{align}
which is purely differential.
\paragraph{$2r-1<0:$}
\[\pi_{0}\left( \frac{\delta H}{\delta L}\right)
=(...)p^{N-1}+...+(...)p^{-m}+...+(...)p^{-m+(2r-1)},\]
hence $L$ is not properly defined and a Dirac reduction is
required for additional lower order terms. The simplest case is
$r=0$ with one-field reduction. Let
\[ \overline{L}=p^{N}+u_{N-1}p^{N-1}+...+u_{1-m}p^{1-m}+u_{-m}p^{-m}+up^{-m-1}.\]
The Dirac reduction with the constraint $u=0$ lead to the second
Poisson map in the form
\begin{align}\label{quadk1b}
&\pi_{0}^{red}\left( \frac{\delta H}{\delta L}\right)=\nonumber\\
&\left\{ \left( L\frac{\delta H}{\delta L}\right)_{\geq
1},L\right\}_{0}+L\left(\left\{L,\frac{\delta H}{\delta
L}\right\}_{0}\right)_{\geq
-1}+\frac{1}{m}\left\{L,\partial_{x}^{-1}res_{0}\left\{L,\frac{\delta
H}{\delta L}\right\}_{0}\right\}_{0},
\end{align}
which is purely differential. This special case was considered
recently in \cite{CT}.

Observing \eqref{poten} for $n\me 1$ one obtains the highest and
lowest order of $\pi_{n}\bra{\var{H}{L}}$ as
\begin{align*}
\pi_{n}\bra{\var{H}{L}} &= \bra{(...)p^{N-1} + lower} + \bra{(...)p^{(n+1)N+2r-2} + lower}\\
&= \bra{higher + (...)p^{-m}} + \bra{higher +
(...)p^{-(n+1)m+2r-1}},
\end{align*}
where $lower$ ($higher$) represents lower (higher) orders. Hence,
the Lax operators of the type \eqref{slaxk1} do not form a proper
Poisson submanifold for the $\pi_n$ with $n\me 1$, apart from the
trivial case of $N=-m=\frac{1-2r}{n}$. Hence, one has to apply
Dirac reduction to restrict the bivectors $\pi_n$ on the
polynomials to the submanifold of the form \eqref{poten}.

\paragraph{The case of $k=2$.}
This has not been considered yet. Let us consider a Lax polynomial
\eqref{laxk2} of the form
\begin{equation}\label{slaxk2}
L= u_N p^N + u_{N-1} p^{N-1} + ... + u_{1-m} p^{1-m} + p^{-m}.
\end{equation}
Then, gradient of functional $H(L)$ is given in the form
\begin{equation}
\var{H}{L} = \var{H}{u_{1-m}} p^{r+m-2} + ... + \var{H}{u_{N-1}}
p^{r-N} + \var{H}{u_{N}} p^{r-N-1}.
\end{equation}
Then by analogous consideration as for $k=1$ or by Theorem
\ref{rel}, for the first Poisson structure with $n=-1$, $L$
defines a proper Poisson submanifold for $N\geq 2r-3\geq -m,$ so
the first Poisson map in this case is
\begin{equation}\label{link2}
\pi _{-1}\left( \frac{\delta H}{\delta L}\right) =\left\{
\left(\frac{\delta H}{\delta L}\right)_{\geq
-r+2},L\right\}_{r}+\left( \left\{ L,\frac{\delta H}{\delta
L}\right\}_{r}\right)_{\geq 2r-2}.
\end{equation}
Otherwise a Dirac reduction is required.

For the second Poisson map with $n=0$, $L$ does not define a
proper Poisson submanifold and again two distinct cases have to be
considered.

\paragraph{$2r-3>0:$}
\[ \pi_{0}\left(\frac{\delta H}{\delta L}\right) =
(...)p^{N+(2r-3)}+...+(...)p^{N}+...+(...)p^{1-m},\]
hence $L$ is not properly defined and a Dirac reduction is
required for additional higher order terms. The simplest case is
$r=2$ with one-field reduction. Let
\[ \overline{L}=up^{N+1}+u_{N}p^{N}+u_{N-1}p^{N-1}+...+u_{1-m}p^{1-m}+p^{-m},\]
then the Dirac reduction with the constraint $u=0$ leads to the
second Poisson map in the form
\begin{align}\label{quadk2a}
&\pi_{0}^{red}\left(\frac{\delta H}{\delta L}\right)=\nonumber\\
&\left\{ \left( L\frac{\delta H}{\delta L}\right) _{\geq
0},L\right\}_{2}+L\left( \left\{ L,\frac{\delta H}{\delta
L}\right\} _{2}\right) _{\geq 2}+\frac{1}{N}\left\{ L,\partial
_{x}^{-1}res_{2}\left\{ L,\frac{\delta H}{\delta L}\right\}
_{2}\right\} _{2},
\end{align}
which is purely differential.

\paragraph{$2r-3<0:$}
\[ \pi _{0}\left(\frac{\delta H}{\delta L}\right)
=(...)p^{N}+...+(...)p^{-m+1}+...+(...)p^{-m+1+(2r-3)},\]
hence $L$ is not properly defined and a Dirac reduction is
required for additional lower order terms. The simplest case is
$r=1$ with one-field reduction. Let
\[ \overline{L}=u_{N}p^{N}+u_{N-1}p^{N-1}+...+u_{2-m}p^{2-m}+u_{1-m}p^{1-m}+up^{-m},\]
then the Dirac reduction with the constraint $u=1$ leads to the
second Poisson map in the form
\begin{align}\label{quadk2b}
&\pi_{0}^{red}\left(\frac{\delta H}{\delta L}\right)=\nonumber\\
&\left\{ \left( L\frac{\delta H}{\delta L}\right) _{\geq
1},L\right\}_{1}+L\left( \left\{ L,\frac{\delta H}{\delta
L}\right\} _{1}\right) _{\geq 0}+\frac{1}{m}\left\{ L,\partial
_{x}^{-1}res_{1}\left\{ L,\frac{\delta H}{\delta L}\right\}
_{1}\right\} _{1},
\end{align}
which is again purely differential.

Now we present one example of three-field Dirac reduction. Let us
consider the case with $r=0$, then
\[ \overline{L}=u_{N}p^{N}+...+u_{1-m}p^{1-m}+up^{-m}+vp^{-m-1}+wp^{-m-2}.\]
The Dirac reduction with constraints $u=1,\ v=w=0$ gives the
following reduced Poisson map
\begin{align}\label{quadk2c}
&\pi_{0}^{red}\left(\frac{\delta H}{\delta L}\right)=\nonumber\\
&\left\{ \left( L\frac{\delta H}{\delta L}\right) _{\geq
2},L\right\}_{0}+L\left( \left\{ L,\frac{\delta H}{\delta
L}\right\}_{0}\right)_{\geq -2}+\left\{
L,A\,p+B+Cp^{-1}\right\}_{0},
\end{align}
where
\begin{align*}
C =&\frac{1}{m}\partial_{x}^{-1}\left(\left\{ L,\frac{\delta H}{\delta L}\right\}_{0}\right)_{-2},\\
B =&\frac{1}{m}\partial_{x}^{-1}\left(\left\{ L,\frac{\delta H}{\delta L}\right\}_{0}\right)_{-1}+\frac{1}{m}u_{-m+1}C,\\
A =&\frac{1}{m}\partial_{x}^{-1}res_{0}\left\{ L,\frac{\delta H}{\delta L}\right\}_{0}+\frac{1}{m^{2}}\partial_{x}^{-1}u_{-m+1}\left( \left\{ L,\frac{\delta H}{\delta L}\right\}_{0}\right)_{-1}\\
&+\frac{1}{m}\left(u_{-m+2}-\frac{1}{2}\frac{m-1}{m}u_{-m+1}^{2}\right)
C+\frac{1}{m}\partial_{x}^{-1}\left(u_{-m+2}-\frac{1}{2}\frac{m-1}{m}
u_{-m+1}^{2}\right) C_{x},
\end{align*}
generally nonlocal.

Then by analogous consideration as for $k=1$ or by Theorem
\ref{rel}, we see that, Lax operators of the form \eqref{slaxk2}
do not form a proper Poisson submanifold for the $\pi_n$ with
$n\me 1$, apart from the trivial case of $N=-m=\frac{3-2r}{n}$.
Hence, one has to apply the Dirac reduction to restrict the
bivectors $\pi_n$ on the polynomials to the submanifold of the
form \eqref{poten}.

Hence we know the Poisson structure for (1+1)-dispersionless
systems constructed from Poisson algebras, and since we are
interested in multi-Hamiltonian systems
\begin{equation}
L_{t_q} = \{(L^q)_{\me -r+k},L\}_r = \pi_{-1}dH_1 = \pi_{0}dH_0 =
\pi_{-1}dH_{-1} = ...\ ,
\end{equation}
we shall now consider the problem of their construction. The
conserved quantities $H_i$ from \eqref{casq} are defined as
follows
\begin{equation}
H_i(L) = \frac{1}{q+i} tr(L^{q+i}) = \frac{1}{q+i} \int_{\Omega}
res_r\bra{L^{q+i}}\ dx.
\end{equation}

\section{A list of some (1+1)-dimensional dispersionless systems}

In this section we will display a list of the simplest nonlinear
dispersionless integrable systems. Calculating the powers
$L^{\frac{n}{N}}$ we consider the Lax hierarchy
\begin{equation}\label{laxn}
L_{t_n} = \pobr{\bra{L^{\frac{n}{N}}}_{\me -r+k},L}_r,\qquad
n=1,2,3,...\ .
\end{equation}
The second hierarchy with powers $L^{\frac{n}{m}}$, can be
obtained by the transformation from Theorem \ref{rel}, which we
leave for the interested reader. In general for simplicity we
present only the bi-Hamiltonian structure. For $k=0$ and $k=1$ the
choice $n=1-r$ will always lead to the dynamics
$(u_i)_{t_{1-r}}=(1-r)(u_i)_x$ for the fields $u_i$ in $L$, so
that we may identify $t_{1-r}=\frac{1}{1-r}x$ in this cases. For
$k=0$ and integer values of $n/N$ the equations become trivial,
because then $(L^{\frac{n}{N}})_{\me 0}=L$. For each choice of
$k=0,1$ or $2$ and $N$ we will exhibit the first nontrivial of the
nonlinear Lax equations \eqref{laxn} associated with a chosen
operator $L$.

\paragraph{The case of $k=0$.}

\begin{example}
Dispersionless Korteweg-de Vries: $k=0,r=0,N=2$.

\rm This is a standard case of the dispersionless Korteweg-de Vries
(dKdV) hierarchy. The Lax operator for the dKdV has the form
\begin{equation}
L = p^2 + u.
\end{equation}
We derive the dKdV equation
\begin{eqnarray}
L_{t_3} = \pobr{\bra{L^{\frac{3}{2}}}_{\me 0},L}_0 \iff u_{t_3} =
\frac{3}{2}uu_x=\pi_{-1}dH_1=\pi_0^{red}dH_0=\pi_1^{red}dH_{-1},
\end{eqnarray}
where we get the Poisson tensors from \eqref{link0} and
\eqref{quadk0}
\begin{equation}
\begin{split}
&\pi_{-1} = 2 \Dx{},\quad \pi_0^{red} = \Dx{}u+u\Dx{},\\
&\pi_1^{red}=\pi_0^{red}\bra{\pi_{-1}}^{-1}\pi_0^{red} =
\Dx{}u^2+u^2\Dx{}-\frac{1}{2}u_x\Dx{-1}u_x,
\end{split}
\end{equation}
and the respective Hamiltonians
\begin{equation}
H_1 = \frac{1}{8} \int_\Omega u^3\ dx,\quad H_0 = \frac{1}{4}
\int_\Omega u^2\ dx,\quad H_{-1} = \int_\Omega u\ dx.
\end{equation}
\end{example}

\begin{example}
Dispersionless Bousinesq: $k=0,r=0,N=3$.

\rm The Lax operator is given by
\begin{equation}
L= p^3 + up + v.
\end{equation}
We derive
\begin{eqnarray}
L_{t_2} = \pobr{\bra{L^{\frac{2}{3}}}_{\me 0},L}_0 \iff
\Matrix{t_2}{u}{v} = \Matrix{}{2v_x}{-\frac{2}{3}uu_x} =
\pi_{-1}dH_1=\pi_0^{red}dH_0.
\end{eqnarray}
Eliminating the field $v$ from this equation we can derive the
(1+1)-dimensional dispersionless Boussinesq equation
\begin{equation}
u_{tt} = - \frac{2}{3}(u^2)_{xx}.
\end{equation}
The respective Poisson tensors are
\begin{equation}
\pi_{-1} = 3
\pmatrix{cc}{0 & \Dx{}\\ \Dx{} & 0},
\quad \pi_0^{red} =
\pmatrix{cc}{\Dx{}u+u\Dx{} & 2\Dx{}v+v\Dx{}\\
\Dx{}v+2v\Dx{} & -\frac{2}{3}u\Dx{}u},
\end{equation}
and the Hamiltonians are given in the following form
\begin{equation}
H_1 = \frac{1}{3} \int_\Omega (v^2-\frac{1}{9}u^3)\ dx,\quad H_0 =
\int_\Omega v\ dx.
\end{equation}
\end{example}

\begin{example}
The three field case: $k=0, r=0, N=4$.

\rm The Lax operator is
\begin{equation}
L= p^4 + up^2 + vp + w,
\end{equation}
then
\begin{align}
&L_{t_2} = \pobr{\bra{L^{\frac{2}{4}}}_{\me 0},L}_0 \iff \nonumber\\
&\Matrixx{t_2}{u}{v}{w} =
\Matrixx{}{2v_x}{-uu_x+2w_x}{-\frac{1}{2}u_xv} =
\pi_{-1}dH_1=\pi_0^{red}dH_0,
\end{align}
where
\begin{equation}
\begin{split}
\pi_{-1} &=
\pmatrix{ccc}{
0& 0& 4\Dx{}\\
0& 4\Dx{}& 0\\
4\Dx{} &0 &\Dx{}u+u\Dx{}
},\\
\pi_0^{red}&=
\pmatrix{ccc}{
\Dx{}u+u\Dx{}& 2\Dx{}v+v\Dx{}& 3\Dx{}w+w\Dx{}\\
\Dx{}v+2v\Dx{}& -u\Dx{}u+2\Dx{}w+2w\Dx{}& -\frac{1}{2}u\Dx{}v\\
\Dx{}w+3w\Dx{}& -\frac{1}{2}v\Dx{}u&
-\frac{3}{4}v\Dx{}v+u\Dx{}w+w\Dx{}u
},
\end{split}
\end{equation}
\begin{equation}
H_1 = \frac{1}{2}\int_\Omega (-\frac{1}{4}u^2v+vw)\ dx,\quad H_0 =
\int_\Omega v\ dx.
\end{equation}
\end{example}

\paragraph{The case of $k=1$.}

\begin{example}
Three field hierarchy: $k=1, r\in \bb{Z}\setminus \{2\}$.

\rm The Lax operator has the form \eqref{laxk1} with $N=2-r,\ m=r+1$
\begin{equation}\label{lax1Z}
L = p^{2-r} + up^{1-r} + vp^{-r} + wp^{-r-1},
\end{equation}
Then we find
\begin{align}\label{lax1Ze}
&L_{t_{2-r}} = \pobr{\bra{L}_{\me -r+1},L}_r \iff \nonumber\\
&\Matrixx{t_{2-r}}{u}{v}{w}=
\Matrixx{}{(2-r)v_x}{ru_xv+(1-r)uv_x+(2-r)w_x}{(1+r)u_xw+(1-r)uw_x},
\end{align}
This Lax operator forms a proper submanifold as regards $\pi_{-1}$
only for $r=0,1$. Otherwise a Dirac reduction is required. Then
for $r=0$
\begin{equation}
\Matrixx{t_2}{u}{v}{w} = \Matrixx{}{2v_x}{uv_x+2w_x}{u_xw+uw_x} =
\pi_{-1}dH_1 = \pi_0^{red}dH_0,
\end{equation}
where
\begin{equation}
\begin{split}
\label{todar0}
\pi_{-1} &=
\pmatrix{ccc}{
0& 0& 2\Dx{}\\
0& 2\Dx{}& u\Dx{}\\
2\Dx{}& \Dx{}u &0
},\\
\pi_0^{red} &=
\pmatrix{ccc}{
6\Dx{}& 4\Dx{}u& 2\Dx{}v\\
4u\Dx{}& 2u\Dx{}u+\Dx{}v+v\Dx{}& u\Dx{}v+2\Dx{}w+w\Dx{}\\
2v\Dx{}& v\Dx{}u+\Dx{}w+2w\Dx{}& u\Dx{}w+w\Dx{}u
},
\end{split}
\end{equation}
\begin{equation}
H_1 = \int_\Omega vw\ dx,\quad H_0 = \int_\Omega w\ dx.
\end{equation}
For $r=1$ we have
\begin{equation}\label{todar1}
\Matrixx{t_1}{u}{v}{w} = \Matrixx{}{v_x}{u_xv+w_x}{2u_xw} =
\pi_{-1}dH_1 = \pi_0^{red}dH_0,
\end{equation}
where
\begin{equation}
\begin{split}
\pi_{-1} &=
\pmatrix{ccc}{
0& \Dx{}v& 2\Dx{}w\\
v\Dx{}& \Dx{}w+w\Dx{}& 0\\
2w\Dx{}& 0& 0
},\\
\pi_0^{red} &=
\pmatrix{ccc}{
\Dx{}v+v\Dx{}& u\Dx{}v+2\Dx{}w+w\Dx{}& 2u\Dx{}w\\
v\Dx{}u+\Dx{}w+2w\Dx{}& 2v\Dx{}v+u\Dx{}w+w\Dx{}u& 4v\Dx{}w\\
2w\Dx{}u& 4w\Dx{}v& 6w\Dx{}w
},
\end{split}
\end{equation}
\begin{equation}
H_1 = \frac{1}{2} \int_\Omega (u^2+2v)\ dx,\quad H_0 = \int_\Omega u\ dx.
\end{equation}
\end{example}

\begin{example}
Dispersionless Toda: $k=1, r\in \bb{Z}\setminus \{2\}$

\rm The first admissible reduction $w=0$ of (\ref{lax1Z}) leads to the
two field Lax operator
\begin{equation}\label{lax1Zi}
L = p^{2-r} + up^{1-r} + vp^{-r}.
\end{equation}
This Lax operator forms a proper submanifold as regards $\pi_{-1}$
only for $r=1$, in other cases a Dirac reduction is required. For
$r=1$ by reduction of \eqref{todar1} we get the dispersionless
Toda equation
\begin{equation}
\Matrix{t_1}{u}{v} = \Matrix{}{v_x}{u_xv} = \pi_{-1}dH_1 =
\pi_0^{red}dH_0,
\end{equation}
where
\begin{equation}
\begin{split}
&\pi_{-1} =
\pmatrix{cc}{
0& \Dx{}v\\
v\Dx{}& 0\\
},\quad
\pi_0^{red} =
\pmatrix{cc}{
\Dx{}v+v\Dx{}& u\Dx{}v\\
v\Dx{}u& 2v\Dx{}v
},\\
&H_1 = \frac{1}{2} \int_\Omega (u^2+2v)\ dx,\quad H_0 =
\int_\Omega u\ dx.
\end{split}
\end{equation}
For $r=0$ we have
\begin{equation}
\Matrix{t_2}{u}{v} = \Matrix{}{2v_x}{uv_x},
\end{equation}
but we lose the bi-Hamiltonian structure since there are no Dirac
reductions with the constraint $w=0$ of \eqref{todar0}.
\end{example}

The next admissible reduction $w=v=0$ of \eqref{lax1Ze} leads to
the noninteresting trivial equation $L_{t_{2-r}}=0$ since
$(L)_{\me -r+1}=L$.

\begin{example}
Three field hierarchy: $k=1, r\in \bb{Z}\setminus \{1\}$.

\rm The Lax operator has the form \eqref{laxk1} with $N=1-r,\ m=r+2$
\begin{equation}\label{lax1a}
L = p^{1-r} + up^{-r} + vp^{-r-1} + wp^{-r-2},
\end{equation}
Then we find
\begin{align}
\label{lax1ae}
&L_{t_{2-r}} = \pobr{\bra{L^{\frac{2-r}{1-r}}}_{\me -r+1},L}_r \iff \nonumber\\
&\Matrixx{t_{2-r}}{u}{v}{w}= \frac{2-r}{1-r}
\Matrixx{}{uu_x+(1-r)v_x}{(1+r)u_xv+(1-r)uv_x+(1-r)w_x}{(2+r)u_xw+(1-r)uw_x},
\end{align}
This Lax operator forms a proper submanifold as regards $\pi_{-1}$
only for $r=0$, in other cases a Dirac reduction is required. Then
for $r=0$ we have
\begin{equation}
\label{benny}
\Matrixx{t_2}{u}{v}{w} =
2\Matrixx{}{uu_x+v_x}{u_xv+uv_x+w_x}{2u_xw+uw_x} = \pi_{-1}dH_1 =
\pi_0^{red}dH_0,
\end{equation}
where
\begin{equation}
\begin{split}
\label{b1}
\pi_{-1} &=
\pmatrix{ccc}{
0& \Dx{}& 0\\
\Dx{}& 0& 0\\
0& 0& \Dx{}w+w\Dx{}
},\\
\pi_0^{red} &=
\pmatrix{ccc}{
\frac{3}{2}\Dx{}& \Dx{}u& \frac{1}{2}\Dx{}v\\
u\Dx{}& \Dx{}v+v\Dx{}& 2\Dx{}w+w\Dx{}\\
\frac{1}{2}v\Dx{}& \Dx{}w+2w\Dx{}&
-\frac{1}{2}v\Dx{}u+u\Dx{}w+w\Dx{}u
},
\end{split}
\end{equation}
\begin{equation}
H_1 = \int_\Omega (u^2v+v^2+2uw)\ dx,\quad H_0 = \int_\Omega (uv+w)\ dx.
\end{equation}
\end{example}

\begin{example}
Benney system: $k=1, r\in \bb{Z}\setminus \{1\}$

\rm The first admissible reduction $w=0$ of (\ref{lax1a}) leads to the
two field Lax operator
\begin{equation}
L = p^{1-r} + up^{-r} + vp^{-r-1}.
\end{equation}
This Lax operator forms a proper submanifold as regards $\pi_{-1}$
only for $r=0$, otherwise a Dirac reduction is required. For $r=0$
by reduction of \eqref{benny} we get the Benney system
\begin{equation}\label{benny0}
\Matrix{t_2}{u}{v} = 2\Matrix{}{uu_x+v_x}{u_xv+uv_x} =
\pi_{-1}dH_1 = \pi_0^{red}dH_0,
\end{equation}
where
\begin{equation}
\begin{split}
&\pi_{-1} =
\pmatrix{cc}{
0& \Dx{}\\
\Dx{}& 0\\
},\quad
\pi_0^{red} =
\pmatrix{cc}{
2\Dx{}& \Dx{}u\\
u\Dx{}& \Dx{}v+v\Dx{}
},\\
&H_1 = \int_\Omega (u^2v+v^2)\ dx,\quad H_0 = \int_\Omega uv\ dx.
\end{split}
\end{equation}
\end{example}

The next admissible reduction $w=v=0$ of \eqref{lax1ae} leads to
\begin{equation}
u_{t_{2-r}} = \frac{2-r}{1-r} uu_x,
\end{equation}
but for $r=0$ we lose the bi-Hamiltonian structure since there are
no Dirac reductions with $w=0$ of \eqref{b1}.

\paragraph{The case of $k=2$.}

\begin{example}
Two field hierarchy: $k=2,r\in \bb{Z}\setminus \{3\}$.

\rm The Lax operator is given by
\begin{equation}
L = u^{3-r}p^{3-r}+vp^{2-r}+p^{1-r},
\end{equation}
then we have
\begin{equation}
\begin{split}
&L_{t_{4-r}} = \pobr{\bra{L^{\frac{4-r}{3-r}}}_{\me -r+2},L}_r \iff\\
&u_{t_{4-r}} = \frac{4-r}{2(3-r)^2} \Bigl(2(3-r)(2r-3)u^{2-r}u_xv+2(3-r)u^{3-r}v_x-(2-r)v^2v_x \Bigr.\\
&\qquad \qquad \qquad \qquad \Bigl.+(2-r)^2(\ln u)_xv^3\Bigr),\\
&v_{t_{4-r}} = \frac{4-r}{2(3-r)^2} \Bigl(2(r-1)(3-r)^2u_x+(1-r)(2-r)(3-r)u^{r-3}u_xv^2\Bigr.\\
&\qquad \qquad \qquad \qquad \Bigl.-2(1-r)(3-r)u^{r-2}vv_x\Bigr).
\end{split}
\end{equation}
For $r=2$ we find
\begin{equation}
\Matrix{t_{2}}{u}{v}= 2 \Matrix{}{u_xv+uv_x}{u_x+vv_x},
\end{equation}
which is again a Benney system with the known bi-Hamiltonian
structure, this time reconstructed from formulae \eqref{link2} and
\eqref{quadk2a}.
\end{example}

\begin{example}
Two field hierarchy: $k=2, r\in \bb{Z}\setminus \{2\}$.

\rm The Lax operator has the form \eqref{laxk2} with $N=2-r,\ m=r$
\begin{equation}
L = u^{2-r}p^{2-r} + vp^{1-r} + p^{-r},
\end{equation}
Then we find
\begin{eqnarray}
L_{t_{2-r}} = \pobr{\bra{L}_{\me -r+2},L}_r \iff
 \Matrix{t_{2-r}}{u}{v}=
\Matrix{}{(r-1)ru_xv+uv_x}{(2-r)ru^{1-r}u_x},
\end{eqnarray}
This Lax operator forms a proper submanifold as regards $\pi_{-1}$
only for $r=1$, otherwise a Dirac reduction is required. Then for
$r=1$
\begin{equation}
\Matrix{t_1}{u}{v}= \Matrix{}{uv_x}{u_x},
\end{equation}
it is again a Toda system with the known bi-Hamiltonian structure,
this time reconstructed from formulae \eqref{link2} and
\eqref{quadk2b}.
\end{example}

\begin{example}
Three field hierarchy: $k=2, r\in \bb{Z}\setminus \{2\}$.

\rm The Lax operator has the form \eqref{laxk1} with $N=2-r,\ m=r+1$
\begin{equation}
L = up^{2-r} + vp^{1-r} + wp^{-r} + p^{-r-1},
\end{equation}
Then we find
\begin{eqnarray}
L_{t_{2-r}} = \pobr{\bra{L}_{\me -r+2},L}_r \iff \Matrixx{t_{2-r}}{u}{v}{w}=
\Matrixx{}{(r-1)u_xv+(2-r)uv_x}{ru_xw+(2-r)uw_x}{(1+r)u_x}.
\end{eqnarray}
This Lax operator forms a proper submanifold as regards $\pi_{-1}$
only for $r=1$, in other cases a Dirac reduction is required. Then
for $r=1$
\begin{equation}
\Matrixx{t_1}{u}{v}{w} = \Matrixx{}{uv_x}{u_xw+uw_x}{2u_x} =
\pi_{-1}dH_1 = \pi_0^{red}dH_0,
\end{equation}
where
\begin{equation}
\begin{split}
\pi_{-1} &=
\pmatrix{ccc}{
0& u\Dx{}& 0\\
\Dx{}u& 0& 0\\
0& 0& 2\Dx{}
},\\
\pi_0^{red} &=
\pmatrix{ccc}{
\frac{3}{2}u\Dx{}u& u\Dx{}v& \frac{1}{2}u\Dx{}w\\
v\Dx{}u& u\Dx{}w+w\Dx{}u& \Dx{}u+2u\Dx{}\\
\frac{1}{2}w\Dx{}u& 2\Dx{}u+u\Dx{}&
-\frac{1}{2}w\Dx{}w+\Dx{}v+v\Dx{}
},
\end{split}
\end{equation}
\begin{equation}
H_1 = \frac{1}{2} \int_\Omega (v^2+2uw)\ dx,\quad H_0 = \int_\Omega v\ dx.
\end{equation}
\end{example}

\begin{example}
Two field hierarchy: $k=2, r\in \bb{Z}\setminus \{1\}$.

\rm The Lax operator has the form \eqref{laxk2} with $N=1-r,\ m=r+1$
\begin{equation}
L = u^{1-r}p^{1-r} + vp^{-r} + p^{-r-1},
\end{equation}
Then we find
\begin{eqnarray}
L_{t_{2-r}} = \pobr{\bra{L^{\frac{2-r}{1-r}}}_{\me -r+2},L}_r \iff \Matrix{t_{2-r}}{u}{v}= \frac{2-r}{1-r}
\Matrix{}{ruu_xv+u^2v_x}{(1-r^2)u^{1-r}u_x}.
\end{eqnarray}
Let us consider the case of $r=0$. To get $\pi_{-1}$ we have to
make a Dirac reduction as the condition $N\geq 2r-3\geq -m$ is
violated. The simplest admissible Lax polynomial has the form
\begin{equation}
\overline{L}=up+v+wp^{-1}+zp^{-2}
\end{equation}
and the Poisson operator reconstructed from \eqref{link2} is
\begin{equation}\label{pos}
\pi_{-1}=\left(
\begin{array}{cccc}
0 & 0 & 0 & 2u\,\partial _{x}-\partial _{x}u \\
0 & 0 & u\,\partial _{x} & -v_{x} \\
0 & \partial _{x}u & 0 & -\partial _{x}w \\
2\partial _{x}u-u\,\partial _{x} & v_{x} & -w\,\partial _{x} &
-\partial
_{x}z-z\,\partial _{x}%
\end{array}
\right) .
\end{equation}
Then, reduction of \eqref{pos} with constraints $z=0,w=1$ gives
\begin{equation}
\pi _{-1}^{red}=\left(
\begin{array}{cc}
0 & u^{2}\partial _{x} \\
\partial _{x}u^{2} & 0%
\end{array}
\right),
\end{equation}
while the second Poisson operator, constructed from
\eqref{quadk2a} takes the form
\begin{equation}
\pi _{0}^{red}=\left(
\begin{array}{cc}
u^{2}\partial _{x}u+u\,\partial _{x}u^{2} &
u^{2}v_{x}+u^{2}\partial _{x}v
\\
-u^{2}v_{x}+v\,\partial _{x}u^{2} & 2u\,\partial _{x}u%
\end{array}
\right) .
\end{equation}
Fortunately, both Poisson operators are again differential. Hence
\begin{equation}
\Matrix{t_2}{u}{v} = 2\Matrix{}{u^2v_x}{uu_x}=\pi_{-1}^{red} dH_1
= \pi_0^{red} dH_0,
\end{equation}
where
\begin{equation}
H_{1}=\int_{\Omega }(u+v^{2})\
dx,\quad H_{0}=\int_{\Omega }v\ dx.
\end{equation}
\end{example}

\begin{example}
Three field hierarchy: $k=2, r\in \bb{Z}\setminus \{1\}$.

\rm The Lax operator has the form \eqref{laxk2} with $N=1-r,\ m=r+2$
\begin{equation}
L = u^{1-r}p^{1-r} + vp^{-r} + wp^{-r-1}+p^{-r-2},
\end{equation}
Then we find
\begin{align}
&L_{t_{2-r}} = \pobr{\bra{L^{\frac{2-r}{1-r}}}_{\me -r+2},L}_r \iff \nonumber\\
&\Matrixx{t_{2-r}}{u}{v}{w}= \frac{2-r}{1-r}
\Matrixx{}{ruu_xv+u^2v_x}{(1-r)u^{1-r}((1+r)u_xw+uw_x)}{(2-r)(1-r)u^{1-r}u_x}.
\end{align}
Let us consider the case for $r=0$. Again condition $N\geq
2r-3\geq -m$ is violated but reducing \eqref{pos} with constraint
$z=1$ we get the first Poisson operator in the form
\begin{align}
&\pi_{-1}^{red}=\nonumber\\
&\pmatrix{ccc}{
\frac{1}{2}u\Dx{}u-\frac{1}{2}u_x\Dx{-1}u_x& \frac{1}{2}uv_x-\frac{1}{2}u_x\Dx{-1}v_x& -\frac{1}{2}uw\Dx{}+\frac{1}{2}u_xw-\frac{1}{2}u_x\Dx{-1}w_x\\
*& -\frac{1}{2}v_x\Dx{-1}v_x& u\Dx{}+\frac{1}{2}v_xw-\frac{1}{2}v_x\Dx{-1}w_x\\
*& *&
\frac{1}{4}w^2\Dx{}+\frac{1}{4}\Dx{}w^2-\frac{1}{2}w_x\Dx{-1}w_x
},
\end{align}
where $\ast $ denotes the elements that make the matrix
skew-adjoint. The second Poisson operator calculated according to
\eqref{quadk2a} is
\begin{align*}
\left( \pi _{0}^{red}\right) _{11} =&\ \frac{1}{4}u^{2}(v-\frac{1}{4}%
w^{2})\partial _{x} +\frac{1}{4}[u(v-\frac{1}{4}w^{2})_{x}-u_{x}(v-\frac{1}{4}w^{2})]\partial
_{x}^{-1}u_{x} \\
&+\frac{1}{4}\partial _{x}\,u^{2}(v-\frac{1}{4}w^{2})+\frac{1}{4}u_{x}\partial _{x}^{-1}[u(v-\frac{1}{4}w^{2})_{x}-u_{x}(v-%
\frac{1}{4}w^{2})],
\end{align*}\begin{align*}
\left( \pi _{0}^{red}\right) _{12} =&\ \frac{1}{4}u^{2}w\partial _{x}+\frac{1}{%
2}u(v-\frac{1}{4}w^{2})v_{x}
+\frac{1}{4}[u(v-\frac{1}{4}w^{2})_{x}-u_{x}(v-\frac{1}{4}w^{2})]\partial
_{x}^{-1}v_{x} \\
&+\frac{1}{4}u_{x}\partial
_{x}^{-1}[uw_{x}-v_{x}(v-\frac{1}{4}w^{2})],
\end{align*}\begin{align*}
\left( \pi _{0}^{red}\right) _{13} =&\ (\frac{3}{2}u^{2}+\frac{1%
}{8}uw^{3})\partial _{x}-\frac{1}{2}u\,u_{x}
+\frac{1}{4}w(2vu_{x}-uv_{x}-\frac{1}{2}w^{2}u_{x}+\frac{1}{2}uww_{x}) \\
&+\frac{1}{4}[u(v-\frac{1}{4}w^{2})_{x}-u_{x}(v-\frac{1}{4}w^{2})]\partial
_{x}^{-1}w_{x} -\frac{1}{2}uvw\partial_x\\
&+\frac{1}{4}u_{x}\partial _{x}^{-1}[2u_{x}-(vw)_{x}+\frac{1}{4}%
(w^{3})_{x}],
\end{align*}\begin{align*}
\left( \pi _{0}^{red}\right) _{22} =\frac{3}{2}u\partial _{x}u+\frac{1}{4}%
[uw_{x}-(v-\frac{1}{4}w^{2})v_{x}]\partial _{x}^{-1}v_{x}
+\frac{1}{4}v_{x}\partial
_{x}^{-1}[uw_{x}-(v-\frac{1}{4}w^{2})v_{x}],
\end{align*}\begin{align*}
\left( \pi _{0}^{red}\right) _{23} =&\ u(v-\frac{1}{4}w^{2})\partial _{x}-%
\frac{1}{2}[u-w(v-\frac{1}{4}w^{2})]v_{x}-\frac{1}{4}uww_{x}\\
&+\frac{1}{4}v_{x}\partial
_{x}^{-1}[2u_{x}-(vw)_{x}+\frac{1}{4}(w^{3})_{x}]
+\frac{1}{4}[uw_{x}-(v-\frac{1}{4}w^{2})v_{x}]\partial
_{x}^{-1}w_{x},
\end{align*}\begin{align*}
\left( \pi _{0}^{red}\right) _{33}=&\ \frac{1}{4}[(vw^{2}-2uw)\partial _{x}+\partial
_{x}(vw^{2}-2uw)]-\frac{1}{16}(w^{4}\partial _{x}+\partial
_{x}w^{4}) \\
&+\frac{1}{4}(2u-vw+\frac{1}{4}w^{3})_{x}\partial _{x}^{-1}w_{x}+\frac{1}{4}%
w_{x}\partial _{x}^{-1}(2u-vw+\frac{1}{4}w^{3})_{x}.
\end{align*}%
Notice that both Poisson structures are nonlocal. Then,
\begin{equation}
\Matrixx{t_2}{u}{v}{w} =
2\Matrixx{}{u^2v_x}{uu_xw+u^2w_x}{2uu_x}=\pi_{-1}^{red}dH_1=\pi_0^{red}dH_0,
\end{equation}
where
\begin{equation}
H_{1}=\int_{\Omega
} (uw^2+v^2w+2uv)\ dx,\quad H_{0}=\int_{\Omega }(u+vw)\ dx.
\end{equation}
\end{example}


\end{document}